\documentclass[aps,preprint]{revtex4}
\usepackage{amsmath}
\usepackage{latexsym}
\usepackage{float}
\usepackage{amssymb}
\usepackage{graphicx}
\usepackage{epsfig}

\begin{document}

\title{Structural and Dynamical Anomalies of a Gaussian Core Fluid: a Mode
Coupling Theory Study}
\author{Lindsey Ann Shall and S. A. Egorov}
\affiliation{Department of Chemistry, University of 
Virginia, Charlottesville, Virginia 22901, USA}

\begin{abstract}

We present a theoretical study of  
transport properties of a liquid comprised of particles
interacting via Gaussian Core pair potential. 
Shear viscosity and self-diffusion coefficient are 
computed on the basis of the mode-coupling theory,
with required structural input obtained from integral equation
theory.  
Both self-diffusion coefficient and viscosity display 
anomalous density dependence, with diffusivity increasing and
viscosity decreasing with density within a particular density range 
along several isotherms below a certain temperature. Our theoretical
results for both transport coefficients are
in good agreement with the simulation data.

\end{abstract}
\maketitle
\newcommand{\nc}{\newcommand}
\nc{\cntr}[1]{\begin{center}{\bf #1}\end{center}}
\nc{\ul}{\underline} 
\nc{\fn}{\footnote}
\nc{\pref}{\protect\ref}
\setlength{\parindent}{.25in}
\nc{\stab}[1]{\begin{tabular}[c]{c}#1\end{tabular}}
\nc{\doeps}[3]{\stab{\setlength{\epsfxsize}{#1}\setlength{\epsfysize}{#2}
\epsffile{#3}}}
\nc{\sq}{^{2}}
\nc{\eps}{\epsilon}
\nc{\sig}{\sigma}
\nc{\lmb}{\lambda}
\nc{\lbr}{\left[}
\nc{\rbr}{\right]}
\nc{\lpar}{\left(} 
\nc{\rpar}{\right)}
\nc{\ra}{\rightarrow}
\nc{\Ra}{\Rightarrow}
\nc{\lra}{\longrightarrow}
\nc{\LRa}{\Longrightarrow}
\nc{\la}{\leftarrow}
\nc{\La}{\Leftarrow}
\nc{\lla}{\longleftarrow}
\nc{\LLa}{\Longleftarrow}
\nc{\EE}[1]{\times 10^{#1}}
\nc{\simleq}{\stackrel{\displaystyle <}{\sim}}
\newcommand{\be}{\begin{equation}}
\newcommand{\ee}{\end{equation}}
\newcommand{\bea}{\begin{eqnarray}}
\newcommand{\eea}{\end{eqnarray}}
\newcommand{\R}{\vec{R}}
\newcommand{\bc}{\begin{center}}
\newcommand{\ec}{\end{center}}
\clearpage

\section{Introduction}
\label{sc1}

Liquids comprised of particles interacting via a Gaussian Core (GC) model pair
potential have been receiving a lot of attention 
recently.\cite{stillinger76,stillinger97,lang00,prestipino05,mausbach06,
mausbach09,may07,ahmed09,krekelberg09,krekelberg09b,pond09,mladek06,louis00,
wensink08,eurich07,archer01} Strong theoretical interest in GC model stems from
the fact that this bounded potential is useful for describing interactions
between inherently penetrable entities, such as polymer 
coils.\cite{louis00,eurich07} As such, both 
thermodynamic\cite{lang00,archer01} and 
dynamic\cite{mausbach06,mausbach09,may07,ahmed09,krekelberg09,krekelberg09b,
pond09} properties of GC fluids and binary mixtures have been extensively
studied. It has been found that their transport coefficients exhibit anomalous
behavior strongly reminiscent of  waterlike model 
systems\cite{hemmer70,sadrlahijany98,jagla99,franzese01,yan05,kumar05,
mittal06b,deoliveira06b,errington06,egorov08c}, with
diffusivity increasing and viscosity decreasing with density over a certain
range of thermodynamic conditions. Furthermore, a strong correlation between
this behavior and structural anomalies quantified via excess entropy has been
demonstrated.\cite{krekelberg09,krekelberg09b,pond09} Both structural and
dynamical anomalous behavior has been rationalized in terms of bounded nature
of GC potential which results in increasing amount of interparticle overlap
with increasing density. In view of the above, it would be of interest to
provide a firm theoretical link between structural and transport properties of
GC fluid on the basis of microscopic statistical mechanical theory. Such a
connection has been recently established for waterlike model 
systems\cite{egorov08c} by combining integral equation theory of structure with
mode-coupling theory (MCT) treatment of dynamics. The goal of the present work
is to develop a similar treatment for the GC model.  
We show that integral equation
theory/MCT combination is indeed capable of capturing anomalous behavior of
transport coefficients of both neat GC fluids and binary mixtures observed in
molecular dynamics (MD) 
simulations.\cite{mausbach06,mausbach09,may07,ahmed09,krekelberg09,
krekelberg09b,pond09}
Our MCT-based microscopic analysis helps
to shed further light on the origin and nature of transport anomalies. In
particular, we are able to rationalize why viscosity anomaly persists over much
more narrow range of densities and temperatures compared to the diffusion
anomaly.  

The remainder of the paper is organized as follows. In
Section~\ref{sc2} we describe the microscopic model and review the MCT
approach employed to calculate the transport coefficients. 
In Section~\ref{sc3} we compare theoretical results with the
simulation data for the shear viscosity and the
self-diffusion coefficient. In Section~\ref{sc4} we conclude. 

\section{Microscopic Model and Theory}
\label{sc2}

We consider a system comprised of spherical particles interacting via
isotropic GC pair potential $\phi(r)$: 
\be
\phi(r)=\epsilon\exp\left[-(r/\sigma)^{2}\right],
\label{phi}
\ee
where the two parameters $\eps$ and $\sig$ characterize the height and the
width of the interaction profile, respectively.
In the previous studies of the GC model, integral equation theory
with the hypernetted chain (HNC) closure was shown to give reliable structural
results for a wide range of densities studied.\cite{lang00} Hence, we employ
the HNC closure throughout this study to compute the radial distribution
function
$g(r)$ for a liquid described by the GC pair potential. The
validity of this approach will be further confirmed by comparing various
structural quantities with the corresponding simulation data. 
 
The main focus of the present work is the calculation of the transport 
properties of GC fluid and analysis of their anomalous behavior. 
We first describe our treatment of the self-diffusion coefficient. 
The latter is obtained from the total time
integral of the time-dependent friction 
$\zeta(t)$:\cite{balucani94,bhattacharyya97}   
\be
D=\frac{k_BT}{m\zeta_{0}},
\label{dc}
\ee
with
\be
\zeta_{0}=\int_{0}^{\infty}dt\zeta(t).
\label{zeta0}
\ee
The MCT result for the time-dependent friction reads:
\be
\zeta(t)=\frac{k_BT\rho}{6\pi^2m}
\int_{0}^{\infty}dk k^4 c(k)^2 F(k,t) F_s(k,t)
\label{zetamct}
\ee
where $m$ is the mass of the fluid particle, $T$ is the temperature, 
$\rho$ is the number density, 
and $c(k)$ is the fluid direct correlation function, which we obtain from the
HNC closure. In the above, $F(k,t)$ is the fluid dynamic structure factor, 
which we compute from the 
continued fraction representation of its Laplace 
transform truncated at the second order:\cite{balucani94,hansen86}    
\be
F(k,z)=
\frac{S(k)}
{z+\frac{\displaystyle{\delta_1(k)}}
{\displaystyle{z+}\frac{\displaystyle{\delta_2(k)}}
{\displaystyle{z+}\displaystyle{\tau^{-1}(k)}},
}}
\label{fkz}
\ee
where $S(k)$ is the fluid static structure factor, and 
$\delta_i(k)$ is the initial time value of the $i^{\mbox{th}}$
order memory function (MF) of $F(k,t)$.
For the parameter 
$\tau^{-1}(k)$ we use the expression due to Lovesey:\cite{lovesey71}  
$\tau^{-1}(k)=2\sqrt{\delta_2(k)/\pi}$. 
The quantities $\delta_1(k)$ and $\delta_2(k)$  can be easily calculated
from the first three short-time expansion coefficients of $F(k,t)$; the
microscopic expressions for the latter are well-known and will not be
reproduced here.\cite{balucani94,hansen86,bansal77} 
Finally, $F_s(k,t)$ is the fluid self-dynamic structure 
factor, for which we have adopted a simple Gaussian 
model:\cite{balucani94,bhattacharyya97}    
\be
F_s(k,t)=\exp\left[\frac{-k_BTk^2}{m\zeta_{0}}
\left(t+\frac{1}{\zeta_{0}}e^{-t\zeta_{0}}-1)\right)\right].
\label{fskt}
\ee 
Given that the self-dynamic structure factor is a function of $\zeta_{0}$,
which, in turn, depends on $F_s(k,t)$ via Eq.~(\ref{zetamct}),  
Eqs.~(\ref{zetamct})-(\ref{fskt}) need to be solved iteratively and
self-consistently. One could use a more accurate model for $F_s(k,t)$
in terms of the velocity time correlation function (TCF) of a tagged fluid
particle,\cite{bhattacharyya98} but our numerical calculations have
shown that this does not change the results for $D$ in a noticeable
way.

We note that the MCT result for the time-dependent friction given by 
Eq.~(\ref{zetamct}) arises from coupling to collective density modes, and, as
such, is only expected to be valid at intermediate to long times, i.e. it 
describes the slowly varying tail of $\zeta(t)$. At short times, one typically
adopts an empirical approach by modeling the initial decay of the time-dependent
friction via some rapidly decaying analytical function (e.g. a Gaussian), which
is constructed in such a way as to preserve the exact short-time behavior
of $\zeta(t)$. \cite{balucani94,hansen86}
In particular, the exact result for the initial time value 
$\zeta(0)$ reads:\cite{balucani94,hansen86}
\be
\zeta(0)=\frac{4\pi\rho}{3m}\int_{0}^{\infty}dr r^2 g(r)
\nabla^2\phi(r).
\label{zetab0}
\ee
At the same time, the zero-time value of the MCT expression for the
time-dependent friction is given by:
\be
\zeta(0)^{\mbox{\tiny{MCT}}}=\frac{k_BT\rho}{6\pi^2m}
\int_{0}^{\infty}dk k^4 c(k)^2 S(k)
\label{zetamct0}
\ee
By comparing the two expressions given by Eqs.~(\ref{zetab0}) and 
(\ref{zetamct0}) and by applying Parseval's theorem, one can see that the two
results are equivalent under the random phase 
approximation (RPA), $c(r)=-\phi(r)/k_BT$. It has been shown in the previous
studies of the GC liquid\cite{louis00} that RPA is quite accurate in
describing its structural properties. As such, MCT method provides
essentially correct zero-time value of the time-dependent friction. Furthermore,
numerical results presented in the next Section demonstrate that the MCT
approach also describes the initial decay of $\zeta(t)$ quite accurately. Hence,
in the present work we do not decompose time-dependent friction into binary and
collective terms, but rather employ Eq.~(\ref{zetamct}) at all times.
As will be seen below, a similar situation occurs in the MCT treatment of the
potential part of the shear stress autocorrelation function (SACF), 
which enters into the calculation of the shear viscosity.  

The microscopic expression for the shear viscosity is given by 
the Green-Kubo formula in terms of the total time integral of
the SACF (i.e. time-dependent shear viscosity $\eta(t)$):\cite{allen87}
\be
\eta=\frac{1}{Vk_BT}\int_{0}^{\infty}dt
\langle J_{xy}(t)J_{xy}(0)\rangle=\int_{0}^{\infty}dt \eta(t).
\label{etat}
\ee
with the off-diagonal components of the stress tensor $J_{xy}$ given
by:
\be
J_{xy}=\sum_{i=1}^{N}mv_{i}^{x}v_{i}^{y}
-\sum_{i=1}^{N}\sum_{j>i}^{N}r_{ij}^{x}
\frac{\partial \phi(r_{ij})}{\partial r_{ij}^{y}}=J_{xy}^{k}+J_{xy}^{p}
\label{jxy}
\ee
In the above, $V$ is the total volume, $N$ is the number of
particles, lower indices $i$ and $j$ label the particles, while upper
indices $x$ and $y$ denote the vector components of the particle
velocities $v_i$ and displacement vector $r_{ij}$ connecting the
particles $i$ and $j$.

According to Eq.~(\ref{jxy}), the off-diagonal elements of the stress tensor
are comprised of kinetic and potential terms, $J_{xy}^{k}$ and $J_{xy}^{p}$,
respectively. As a result, the SACF splits into three individual parts: 
potential-potential, kinetic-kinetic, and mixed kinetic-potential contributions:
\be
\eta(t)=\eta_{pp}(t)+\eta_{kk}(t)+2\eta_{kp}(t)=\frac{1}{Vk_BT}
\left\{\langle J_{xy}^{p}(t)J_{xy}^{p}(0)\rangle +
\langle J_{xy}^{k}(t)J_{xy}^{k}(0)\rangle +
2\langle J_{xy}^{k}(t)J_{xy}^{p}(0)\rangle\right\},
\label{etasum}
\ee
with the shear viscosity coefficient given by the sum of the total time
integrals of the three terms above:\\
 $\eta=\eta_p+\eta_k+2\eta_{kp}$, where 
$\eta_p=\int_{0}^{\infty}dt\eta_{pp}(t)$ etc.
We now discuss the calculation of each of these terms in turn. 

As in the case of the time-dependent friction, the MCT treatment of the 
potential
term, $\eta_{pp}(t)$, involves coupling to collective density modes, and the
corresponding result reads:\cite{balucani94,hansen86}
\be
\eta_{pp}(t)=\frac{k_BT}{60\pi^2}
\int_{0}^{\infty}dk k^4 \left[\frac{S^{\prime}(k)}{S(k)}\right]^2
\left\{\left[\frac{F(k,t)}{S(k)}\right]^2\right\},
\label{etappt}
\ee
where prime denotes differentiation with respect to the argument. 
In general, the above expression is used at intermediate and long
times, while short-time behavior of $\eta_{pp}(t)$ is modeled
phenomenologically, based on its short-time expansion coefficients. The exact
result for the initial time value of $\eta_{pp}(t)$
is given by:\cite{balucani94,hansen86}
\be
\eta_{pp}(0)=\frac{2\pi}{15}\rho^2\int_{0}^{\infty}dr g(r)\frac{d}{dr}
\left[r^4\frac{d\phi(r)}{dr}\right],
\label{etapp0}
\ee
while the corresponding MCT result can be obtained 
from Eq.~(\ref{etappt}):
\be
\eta_{pp}(0)^{\mbox{\tiny{MCT}}}=\frac{k_BT}{60\pi^2}
\int_{0}^{\infty}dk k^4 \left[\frac{S^{\prime}(k)}{S(k)}\right]^2.
\label{etamct0}
\ee
By comparing the two expressions given by Eqs.~(\ref{etapp0}) and 
(\ref{etamct0}) and by applying Parseval's theorem, one can see that the two
results agree when RPA holds. As will be seen from
our numerical results presented
in the next Section, not only zero-time value of $\eta_{pp}(t)$, but also  
its initial rapid decay is well reproduced by the MCT.
Hence, we use Eq.~(\ref{etappt}) to model the potential part of the
time-dependent shear viscosity at all times. 

Next, the MCT expression for the mixed kinetic-potential term is given 
by:\cite{franosch98}
\be
\eta_{kp}(t)^{\mbox{\tiny{MCT}}}=-\frac{m}{15\pi^2}
\int_{0}^{\infty}dk k \frac{S^{\prime}(k)}{S(k)^2}
\left[\frac{\partial F(k,t)}{\partial t}\right]^2.
\label{etakpt}
\ee
In this case, the initial-time value $\eta_{kp}(0)=0$, while the first
(linear) term in the short-time expansion of $\eta_{kp}(t)^{\mbox{\tiny{MCT}}}$
agrees (under the RPA) with the corresponding exact expansion 
coefficient.\cite{sharma95} Once
again, we employ the MCT expression to describe $\eta_{kp}(t)$ at all times.

Finally, the MCT result for the kinetic contribution to the time-dependent
friction reads:\cite{franosch98}
\be
\eta_{kk}(t)^{\mbox{\tiny{MCT}}}=\frac{m^2}{5\pi^2k_BT}
\int_{0}^{k_{max}}dk k^2 \left[\frac{7}{6}C_{tt}^{2}(k,t)+
\frac{1}{3}C_{ll}^{2}(k,t)+C_{tt}(k,t)C_{ll}(k,t)\right].
\label{etakkt}
\ee
The upper cutoff on the wavevector integral, 
$k_{max}=(6\pi^2\rho)^{1/3}$, is determined by the requirement\cite{franosch98} 
that the initial time value $\eta_{kk}(0)^{\mbox{\tiny{MCT}}}$ coincides with
the exact value $\eta_{kk}(0)=\rho k_BT$.
In the above, $C_{ll}(k,t)=-\ddot{F}(k,t)/k^2$ is the longitudinal
current TCF, and $C_{tt}(k,t)$ is the transverse 
current TCF. 
The Laplace transform of the latter (truncated at the second order) is given 
by:\cite{hansen86,balucani94} 
\be
C_{tt}(k,z)=
\frac{k_BT/m}
{z+\frac{\displaystyle{\delta_{1t}(k)}}
{\displaystyle{z+}\frac{\displaystyle{\delta_{2t}(k)}}
{\displaystyle{z+}\displaystyle{\tau^{-1}_{t}(k)}},
}}
\label{ctkz}
\ee
where $\delta_{it}(k)$ is the initial time value of the $i^{\mbox{th}}$
order MF of $C_{tt}(k,t)$.
The quantities $\delta_{1t}(k)$ and $\delta_{2t}(k)$  can be 
obtained from the first three short-time expansion coefficients of
$C_{tt}(k,t)$, which are well-known.\cite{bansal77}

In analogy to the continued fraction representation of $F(k,t)$, 
parameter $\tau^{-1}_{t}(k)$ is taken to be proportional to 
$\sqrt{\delta_{2t}(k)}$: $\tau^{-1}_{t}(k)=\xi\sqrt{\delta_{2t}(k)}$, with 
the proportionality factor $\xi$ determined by requiring that 
$C_{tt}(k,t)$ behaves correctly in the hydrodynamic 
limit:\cite{hansen86,balucani94,bansal77c} 
\be
[C_{t}(k,t)]_{\mbox{\tiny{hyd}}}=\frac{k_BT}{m}
\exp\left[-\frac{k^2\eta t}{\rho m}\right],
\label{ctkthyd}
\ee
where $\eta$ is the shear viscosity coefficient. The above
requirement yields:
\be
\xi=\frac{\eta}{\rho m}\lim_{k\rightarrow 0}\frac{\sqrt{\delta_{2t}(k)}}
{\delta_{1t}(k)}k^2
\label{ksi}
\ee
  
Thus, the calculation of the transverse current TCF 
requires the knowledge of the shear viscosity coefficient (which is needed to
compute parameter $\xi$ above), 
while $\eta$ itself depends on $C_{tt}(k,t)$ via the
kinetic part of the time-dependent shear viscosity. 
Hence, the calculation of the transverse current TCF and 
the viscosity coefficient
needs to be performed iteratively, in analogy to the calculation of the
self-diffusion coefficient and time-dependent friction described earlier.
 
In order to assess the accuracy of the MCT approach described above,
in the next section we compare our theoretical results with the
MD data\cite{mausbach06,may07,mausbach09,krekelberg09,krekelberg09b,pond09}  
for self-diffusion and shear viscosity coefficients of GC fluid. 

\section{Results}
\label{sc3}

Our results will be presented in terms of dimensionless
density and temperature defined by: $\rho^{\ast}=\rho\sigma^{3}$ and
$T^{\ast}=k_BT/\epsilon$.
We also define dimensionless variables for time, self-diffusion coefficient,
and shear viscosity coefficient as follows: 
$t^{\ast}=t(\epsilon/m\sigma^{2})^{1/2}$,  
$D^{\ast}=D(m/\epsilon\sigma^{2})^{1/2}$, and 
$\eta^*=\eta\sigma^{2}/\sqrt{m\epsilon}$.

\begin{figure}
\includegraphics[width=10cm,angle=0]{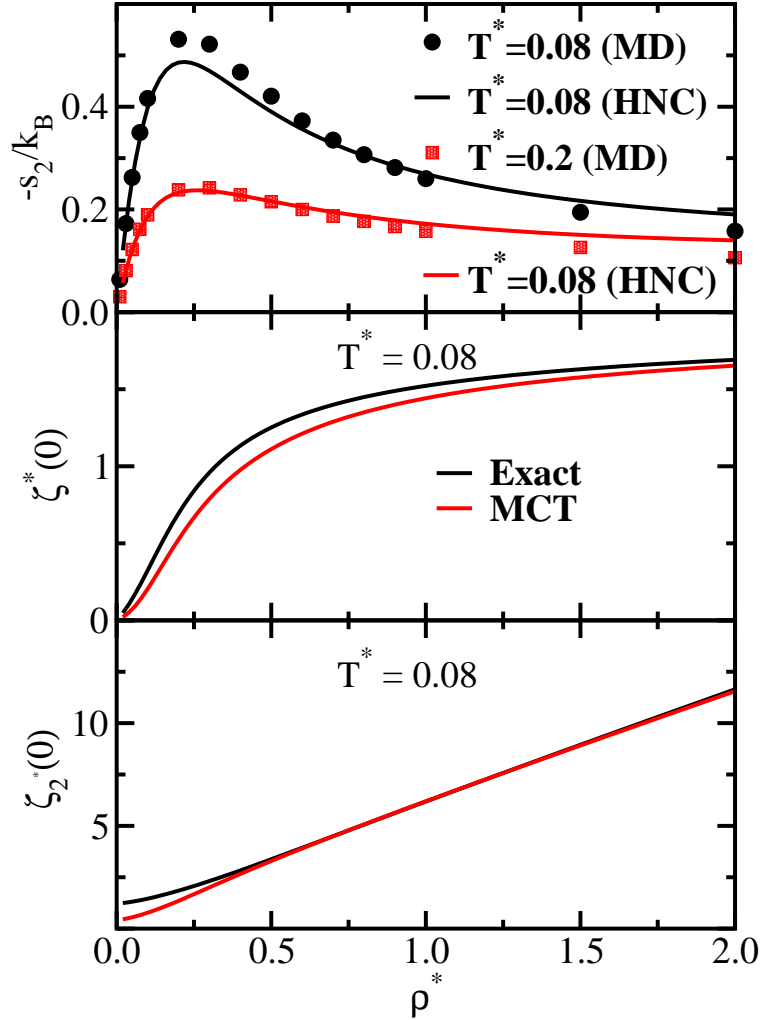}
\caption{Upper panel: Simulation and theoretical results for the negative 
two-body contribution to the excess entropy as a function of fluid density
along two isotherms.
Middle panel: Initial value of the time-dependent friction as given by exact
expression (Eq.~(\ref{zetab0})) and the MCT approach (Eq.~(\ref{zetamct0})). 
Lower panel: Short-time expansion coefficient of order $t^2$ of the normalized
time-dependent friction from exact and MCT approaches.} 
\label{figent} 
\end{figure}

Before presenting and discussing our results for transport coefficients, we
ascertain the accuracy of the HNC closure in calculating structural properties
of GC fluid. This is of particular importance because earlier
studies have revealed a deep connection between transport coefficients and
structural properties, 
such as excess entropy.\cite{krekelberg09,krekelberg09b,pond09}
To this end, we compute from integral equation theory with HNC closure $g(r)$
for the GC fluid, from which one can readily obtain the 
two-body contribution to the excess entropy defined as follows:
\be
s_2=-2\pi\rho\int_{0}^{\infty}dr r^2
\left[g(r)\ln g(r)-(g(r)-1)\right]
\label{s2}
\ee
MD data\cite{krekelberg09} and our HNC results for $-s_2$ are 
presented in the upper panel of Fig.~\ref{figent} 
as a function of fluid density
along two isotherms: $T^*=0.08$ and $T^*=0.2$. In general, theory is in good
agreement with simulations except in the high density range, where it
overestimates $-s_2$ somewhat. Both MD and HNC results show that the
negative two-body excess entropy initially increases with density 
(up to $\rho^*\sim$ 0.25), reflecting increasing structural 
order of the fluid upon its compression. Concomitantly, radial distribution
function (not shown) progressively becomes more structured, with the height of
its first peak increasing and the second solvation shell gradually 
developing.\cite{mausbach06,krekelberg09} 
In simple fluids, whose interaction potential contains a steeply
repulsive core, this familiar type of behavior typically persists throughout the
entire liquid density range. By contrast, GC fluid displays
qualitatively different behavior at higher densities, whereby its structural
order (as measured by $-s_2$) decreases upon further compression, which
manifests itself in flattening of solvation shells in
$g(r)$.\cite{mausbach06,krekelberg09}  This is a consequence of the bounded
nature of the GC potential, which allows interparticle overlap. The
latter becomes more prominent at higher densities and/or temperatures,
ultimately resulting in a high density ideal gas-like
structure.\cite{stillinger76,lang00} 

In the high-density regime, GC system
behaves as a weakly correlated ``mean field fluid'',\cite{louis00} whose direct
correlation function is adequately described by RPA approach. This fact can be
illustrated by comparing the results for the initial time value of the
time-dependent friction given by Eqs.~(\ref{zetab0}) (exact) and 
(\ref{zetamct0}) (MCT), which would yield identical results if RPA were exact.
The results from the two methods for the dimensionless zero-time value 
$\zeta^{*}(0)=\zeta(0)m\sigma^2/\epsilon$ are shown in the
middle panel of Fig.~\ref{figent} as a function of fluid density along the
isotherm $T^*=0.08$.
Indeed, the two approaches give quite similar
values, and the difference between them diminishes with increasing density.  
In addition, we have applied the same two methods to compute the $t^2$
short-time expansion coefficient of the (normalized) time dependent friction, 
$\zeta_2(0)=-\frac{d^2\zeta(t)}{dt^2}|_{t=0}/\zeta(0)$.
The corresponding dimensionless result, 
$\zeta_{2}^{*}(0)=\zeta_{2}(0)m\sigma^2/\epsilon$,  
is presented in the lower panel of Fig.~\ref{figent}. One sees that the two
values are remarkably close numerically (except in the low density regime),
even though the two expressions (exact and MCT) do not become equivalent under
the RPA in this case. 

By comparing the density behavior of  $\zeta(0)$ and $\zeta_2(0)$ shown in
Fig.~\ref{figent}, one observes that the latter increases nearly linearly with
density throughout the entire range, 
while for the former fast increase at low densities is followed by slower
growth at intermediate $\rho$ and nearly flat behavior in the high-density
regime.  These results signify that the initial decay rate of the time-dependent
friction grows continuously with density, while the growth of its zero-time
value with $\rho$ gradually slows down and nearly saturates. 

The above observations can be used to rationalize the density behavior of 
$C_v(t)$, the normalized  velocity TCF of a tagged
GC particle  defined by:
\be
C_v(t)=\frac{m}{k_BT}\langle v_{0}^{x}(t)v_{0}^{x}(0)\rangle,
\label{cvt}
\ee
where $v_{0}^{x}$ is the $x$-component of the tagged particle velocity, and 
$\langle\cdots\rangle$ denotes the Boltzmann equilibrium average.
The Laplace transform of this TCF, $\tilde{C_v}(z)$, can be
related to the Laplace transform of the time dependent friction as 
follows:\cite{hansen86,balucani94} 
\be
\tilde{C_v}(z)=\frac{1}{z+\tilde{\zeta}(z)}.
\label{cvz}
\ee

\begin{figure}
\includegraphics[width=10cm,angle=0]{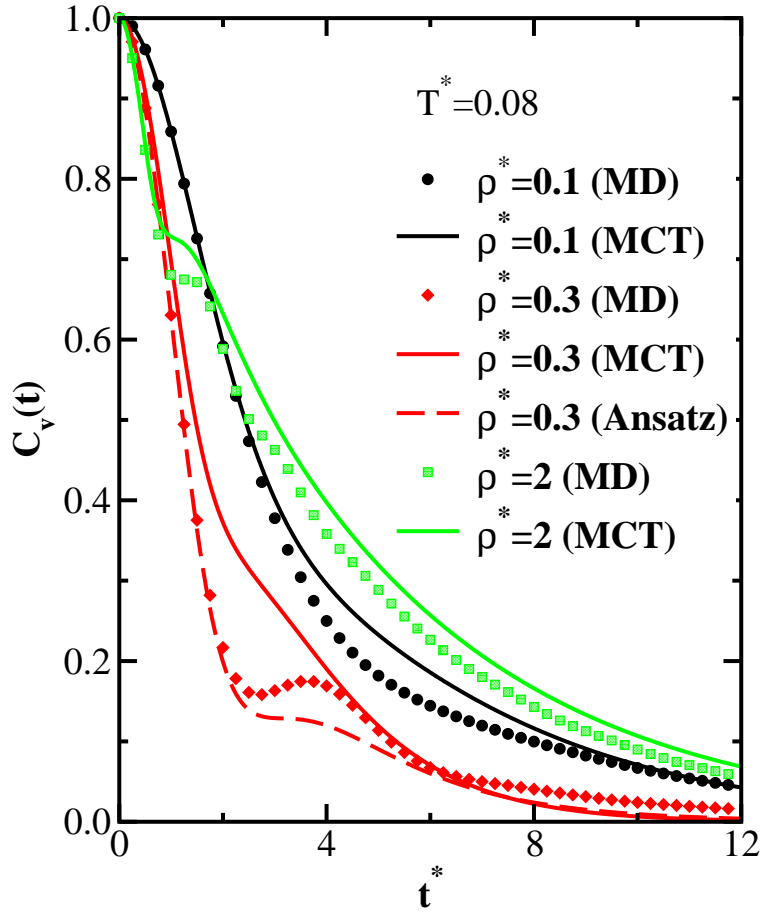}
\caption{Simulation and theoretical results for the normalized
velocity TCF of a tagged GC particle at $T^*=0.08$ at three different 
densities.}
\label{figcvt}
\end{figure}

Simulation\cite{mausbach06} and MCT results for $C_v(t)$ at three different
densities along the isotherm $T^*=0.08$ are shown in Fig.~\ref{figcvt}. 
The most notable feature is biphasic decay of the TCF: except
at the lowest density shown, rapid initial decay of $C_v(t)$ is followed by a
``bump'' at intermediate times and a slowly decaying long-time tail, whose 
amplitude increases markedly with density. The initial decay of the correlation
function is governed by its $t^2$ short-time expansion coefficient, or,
equivalently, by the initial time value of its MF, i.e. time
dependent friction.  As can be seen from the middle panel of Fig.~\ref{figent},
this value initially increases with density, and then gradually saturates.
Accordingly, in going from $\rho^*$=0.1 to $\rho^*$=0.3 the short-time decay of
$C_v(t)$ becomes considerably faster, while for densities beyond $\rho^*$=0.3
it stays essentially unchanged. At the same time, as follows from the lower
panel of Fig.~\ref{figent}, the short-time decay of $\zeta(t)$ itself becomes
faster
with $\rho$ throughout the entire density range. Faster decaying MF
corresponds to slower decaying TCF, which indeed
manifests itself in the amplitude of the slowly decaying tail of $C_v(t)$
growing with density. 

In comparing theoretical results with the simulation data presented in
Fig.~\ref{figcvt}, one sees that the MCT approach reproduces biphasic decay of
the velocity TCF quite well, with the only significant
discrepancy observed at intermediate times for $\rho^*$=0.3. In addition to the
MCT method, we have also explored an alternative approach to constructing 
$C_v(t)$.\cite{sharma95,egorov96} This semi-phenomenological approach is based
on the short-time expansion of the TCF and is frequently
employed in computing transport coefficients via Green-Kubo
relation.\cite{sharma95,egorov96} In particular, one empirical form that is
frequently used to model the normalized TCF is written as follows:
$C_v(t)=\cos(bt)/\cosh(at)$, where parameters $a$ and $b$ are chosen in order to
reproduce the exact short-time behavior of $C_v(t)$ up to the term of order
$t^4$.\cite{sharma95,egorov96} This ansatz has been successfully applied to
calculate diffusion coefficient and other transport properties of simple atomic
liquids, whose interaction potential contains a steeply repulsive term at short
separations.\cite{sharma95,egorov96} 
However, inspection of Fig.~\ref{figcvt} reveals that this
particular form would not be appropriate in the present case, as it would not
be able to capture biphasic decay of the TCF. Instead, we
have attempted to model the time-dependent friction via the above
phenomenological form:
\be
\zeta(t)=\zeta(0)\frac{\cos(bt)}{\cosh(at)}
\label{zetans}
\ee
Such approach requires the knowledge of the short-time
expansion coefficients of $\zeta(t)$ up to the order of $t^4$, or,
equivalently, the expansion coefficients of $C_v(t)$ up to the order of $t^6$.
The latter are well-known and will not be reproduced
here.\cite{tankeshwar87}  

For the density value of $\rho^*$=0.3, we display in Fig.~\ref{figcvt},
alongside with the MCT result, $C_v(t)$ obtained from the time-dependent
friction constructed via Eq.~(\ref{zetans}). One sees that this phenomenological
approach agrees with the MD data even better than the MCT method, both at short
(by construction) and at intermediate times. The performance of ansatz at other
densities is equally good and is not shown to avoid overcrowding the graph. 

As one last remark concerning velocity TCF, we note that
its pronounced biphasic behavior would seemingly suggest applying MCT approach
either to $C_v(t)$ itself,\cite{gaskell78a} or to its 
{\em second-order} MF.\cite{bosse78,egorov03b}
We have carried out such calculations, and found, somewhat surprisingly, that
the results were less accurate compared to the approach based on the
first-order MF outlined above. 

\begin{figure}
\includegraphics[width=10cm,angle=0]{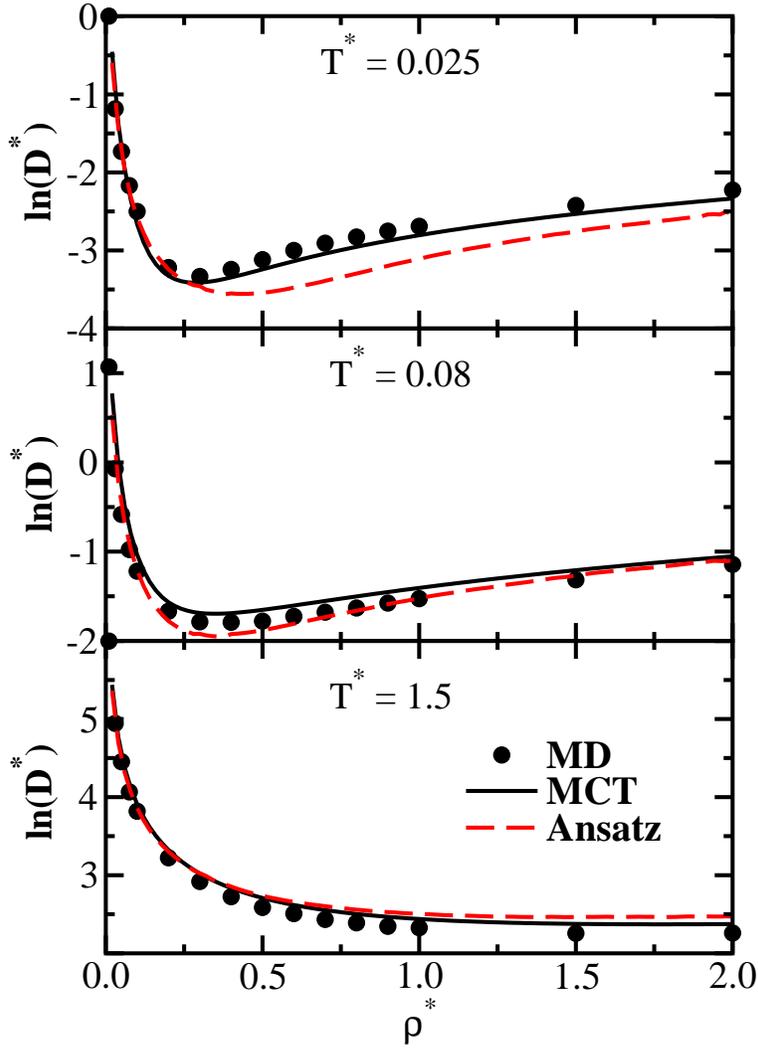}
\vspace{0.8cm} 
\caption{Simulation and theoretical results for the self-diffusion 
coefficient of a tagged GC particle as a function of fluid density along three
isotherms.}
\label{figdiff}
\end{figure}

We next compute the self-diffusion coefficient of GC fluid via Eq.~(\ref{dc}),
with time-dependent friction obtained both from MCT 
(via Eq.~(\ref{zetamct})) and from ansatz (via Eq.~(\ref{zetans})). 
Our results for $D^*$ as a function of GC fluid density, together with MD 
data,\cite{mausbach06,krekelberg09} are shown in Fig.~(\ref{figdiff}). 
MCT is generally in good agreement with the simulation, except that it
overestimates the value of $D$ at intermediate densities at $T^*=0.08$, as one
could already expect from comparing MCT and MD results for the velocity TCF
presented in Fig.~(\ref{figcvt}). By contrast, ansatz systematically
under-predicts $D$ at low and intermediate temperatures. Overall, both methods
correctly reproduce the trends in the density dependence of the self-diffusion
coefficient observed in the simulations.  Specifically, at low and intermediate
temperatures, $D$ initially decreases with $\rho$, passes through a minimum (at
about the same density, for a given temperature, where $-s_2$ passes through a
maximum), and then keeps increasing throughout the remaining density range. For
the highest isotherm shown ($T^*$=1.5), the minimum in $D$ is barely
perceptible, and
at still higher temperatures it disappears altogether (likewise for the maximum
in $-s_2$). As has been already
pointed out,\cite{krekelberg09} a strong correlation between anomalous density
dependencies of $-s_2$ and $D$ points to the structural origin of the transport
anomalies of GC fluid. This conclusion is further re-inforced by the
observation that the density anomaly in $D$ is successfully captured 
by the ansatz expression for time dependent friction, which is constructed 
exclusively from 
the short-time expansion coefficients of $\zeta(t)$. The latter
are simply equilibrium averages of certain functions of the interaction
potential and its derivatives, i.e. purely structural properties.
Some dynamical information does enter the MCT approach, e.g. via the dynamic
structure factor, but, again, $F(k,t)$ was obtained from its continued fraction
representation (Eq.~(\ref{fkz}), which is completely determined by the
short-time expansion coefficients of dynamic structure factor.  

Gradual disappearance of the anomalous density behavior of self diffusion
coefficient at higher temperatures can also be understood in terms of
density dependence of short-time expansion coefficients. In particular, for
$T^*>$1.5, the initial value of time dependent friction no longer saturates at
high densities, but keeps increasing throughout the entire density range (not
shown). As a result, the total time integral of $\zeta(t)$ keeps increasing,
meaning that $D$ decreases monotonically with $\rho$. 
 
Returning to low and intermediate temperatures, 
the anomalous increase of self-diffusion coefficient with density 
(and decrease of structural order) is by no
means unique to GC fluid; this behavior has been observed for several
waterlike model 
systems,\cite{hemmer70,sadrlahijany98,jagla99,franzese01,yan05,kumar05,
mittal06b,deoliveira06b,errington06,egorov08c} 
as well as for colloidal
systems with short-range attractive interactions.\cite{mittal06}
An important difference however, is that for these systems the anomalous 
structural and dynamical behavior is observed for a limited density range only,
after which the system reverts to normal behavior. This is due to the fact that
all these model potentials contain a short-range steeply repulsive term, which
manifests itself in both structure and dynamics at sufficiently high densities.
By contrast, GC potential is bounded, and for this model anomalous behavior
persists till the highest density studied. 

\begin{figure}
\includegraphics[width=10cm,angle=0]{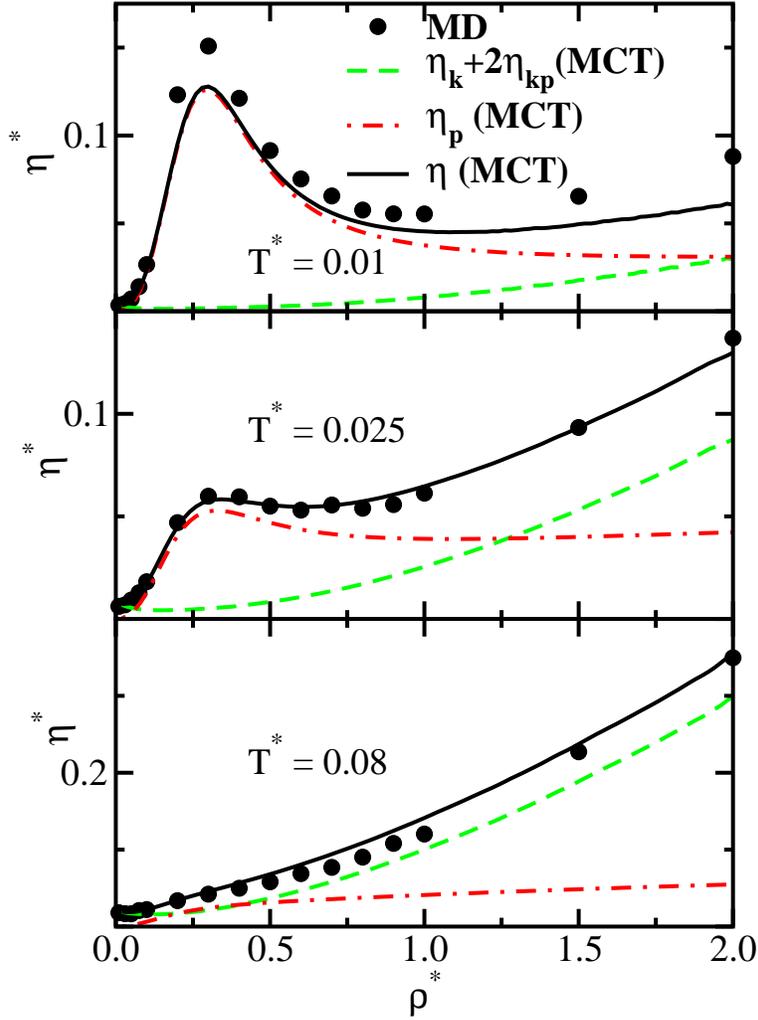}
\vspace{0.8cm} 
\caption{Simulation and theoretical results for the shear viscosity  
of GC fluid as a function of fluid density along three isotherms.}
\label{figetarho}
\end{figure}

We next turn to the discussion of shear viscosity coefficient. 
Simulation\cite{may07,mausbach09} 
and theoretical results for the shear viscosity  
of GC fluid as a function of fluid density along three isotherms are shown in
Fig.~\ref{figetarho}. Also presented are the MCT results for the components of
$\eta$ arising from the potential term, $\eta_p$, and from the sum of kinetic
and mixed kinetic-potential terms, $\eta_k+2\eta_{kp}$. At the lowest
temperature studied, $T^*=0.01$, the shear viscosity coefficient displays a
pronounced anomalous density behavior. Initially, $\eta$ increases rapidly with
$\rho$, passes through a maximum around $\rho^*\sim$0.25, exhibits anomalous
decrease with density until about $\rho^*\sim$1, and then reverts to normal
behavior, i.e. increases with density. This is different from the diffusion
anomaly, where $D$, after passing through a minimum in the low-density region,
keeps increasing with $\rho$ throughout the entire density range studied. In
addition, the temperature range where viscosity anomaly is observed is
substantially more narrow compared to the diffusion anomaly. Thus, along the
isotherm $T^*=0.025$, the minimum and subsequent maximum in $\eta$ are barely
perceptible, both in MD and MCT results, while at still higher temperatures
shear viscosity coefficient increases monotonically with density. 

\begin{figure}
\includegraphics[width=10cm,angle=0]{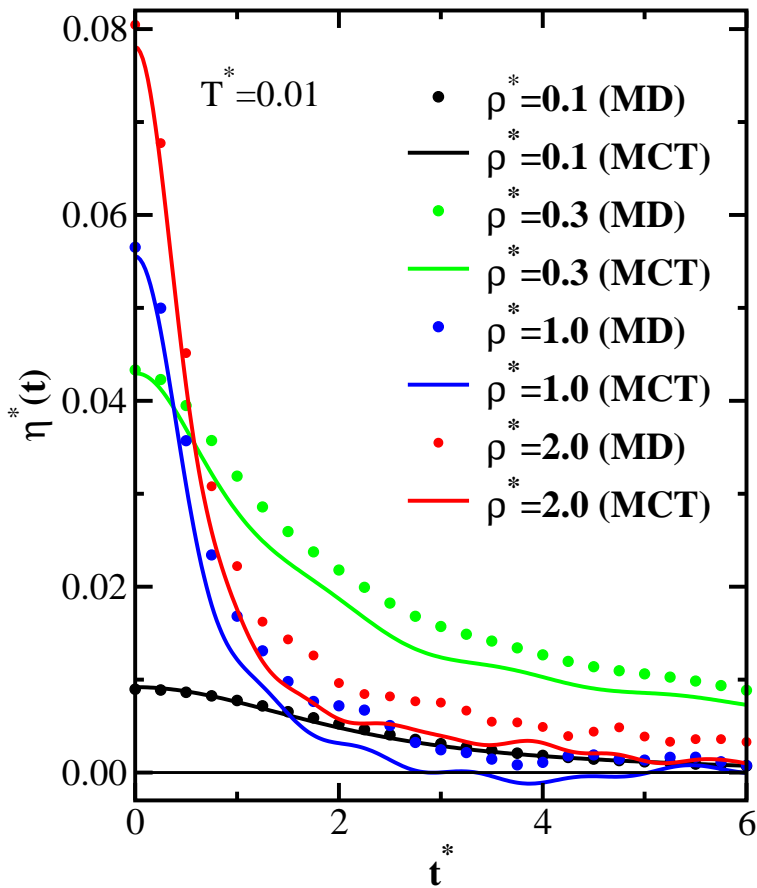}
\vspace{0.8cm} 
\caption{Simulation and theoretical results for the time-dependent shear 
viscosity of GC fluid at $T^*$=0.08 for four different densities.}
\label{figetat}
\end{figure}

If one considers the decomposition of $\eta$ into its components, one observes
that the anomalous behavior at the lowest temperature is dominated by the
potential contribution. Furthermore, after passing through a maximum around
$\rho^*\sim$0.25, $\eta_p$ keeps decreasing with density for all remaining
values of $\rho$. By contrast, $\eta_k+2\eta_{kp}$ term is monotonically
increasing with density. While its magnitude is negligible at low
and intermediate densities, it becomes comparable to $\eta_p$ in the high
density region, which produces a minimum in total $\eta$. As the temperature is
increased, the relative importance of $\eta_k+2\eta_{kp}$ term also increases,
and eventually its contribution to $\eta$ becomes dominant, which results in a
monotonic increase of the shear viscosity coefficient with density. 

\begin{figure}
\includegraphics[width=10cm,angle=0]{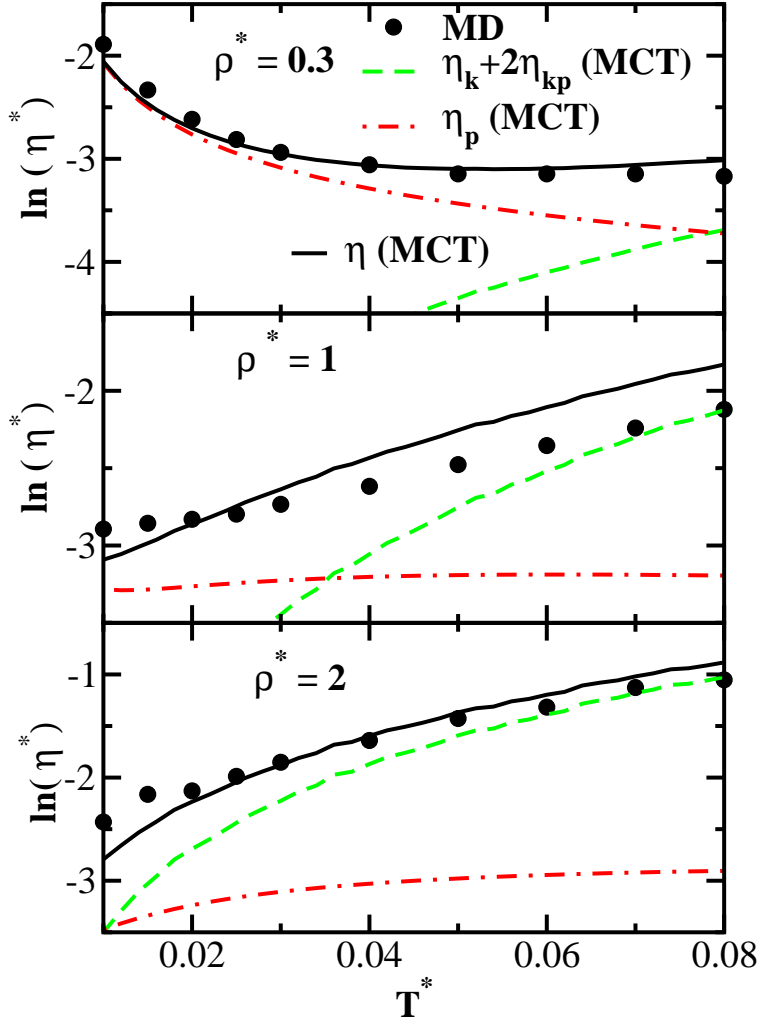}
\vspace{0.8cm} 
\caption{Simulation and theoretical results for the shear 
viscosity of GC fluid as a function of fluid temperature along three isochores.}
\label{figetatmp}
\end{figure}

Some further light on the origins of the anomalous density behavior of
viscosity at low temperatures can be shed by considering the time dependent 
viscosity given by Eq.~(\ref{etasum}). The corresponding simulation and
theoretical results  are shown in Fig.~\ref{etat} for four different
densities along the isotherm $T^*$=0.01. 
One notices that in going from $\rho^*=0.1$ to
$\rho^*=0.3$, the initial time value $\eta(0)$ increases dramatically, while the
decay rate of $\eta(t)$ grows only slightly. As a result, $\eta$, which is
given by the total time integral of $\eta(t)$, grows significantly with density
in this range. The situation is reversed in the intermediate density range,
from $\rho^*$=0.3 to $\rho^*$=1. Here the initial time value grows weakly,
while the decay of $\eta(t)$ becomes much faster, thereby producing anomalous
density behavior of viscosity in this range. Finally, for densities higher than
 $\rho^*$=1, the initial time value keeps increasing gradually, the initial
decay rate does not change significantly, and the amplitude of the long-time
tail of $\eta(t)$ due to mode-coupling effect grows steadily. Hence, the total
time integral of $\eta(t)$ grows with $\rho$, and the shear viscosity
coefficient reverts to normal density behavior. In terms of comparison between
theory and simulation, we note that MCT underestimates the amplitude of the
long-time tail somewhat, which results in the under-prediction of $\eta$ seen in
Fig.~\ref{figetarho}. 

In addition to the density behavior of shear viscosity at a given temperature,
it is also of interest to analyze its temperature behavior at a given
density. To this end, we present in Fig.~\ref{figetatmp} 
simulation\cite{may07,mausbach09} and theoretical results for the shear 
viscosity of GC fluid as a function of fluid temperature along three isochores.
For the isochore $\rho^*$=0.3 (where a maximum in $\eta$ as a function of
$\rho$ is observed), viscosity decreases with temperature at low and
intermediate densities. The initial sharp decrease in this $T$ range is
associated with moving away from the liquid-solid phase
boundary,\cite{may07,mausbach09} and the behavior of $\eta$ is dominated by the
potential term. 
At higher temperatures, $\eta_k+2\eta_{kp}$ term, which is monotonically
increasing with $T$, becomes comparable in magnitude. As a result, the shear
viscosity coefficient passes through a minimum and then grows with temperature. 

As the density of GC fluid grows, its freezing temperature drops.\cite{lang00}
As a result, the initial drop of $\eta$ with $T$ in the studied temperature
range disappears, and $\eta$ increases monotonically with temperature both for
$\rho^*$=1 and for $\rho^*$=2. In the former case, the contributions of kinetic
and potential terms to $\eta$ are comparable, while in the latter case the
kinetic term is dominant for all temperatures except for the lowest one. 

\begin{figure}
\includegraphics[width=10cm,angle=0]{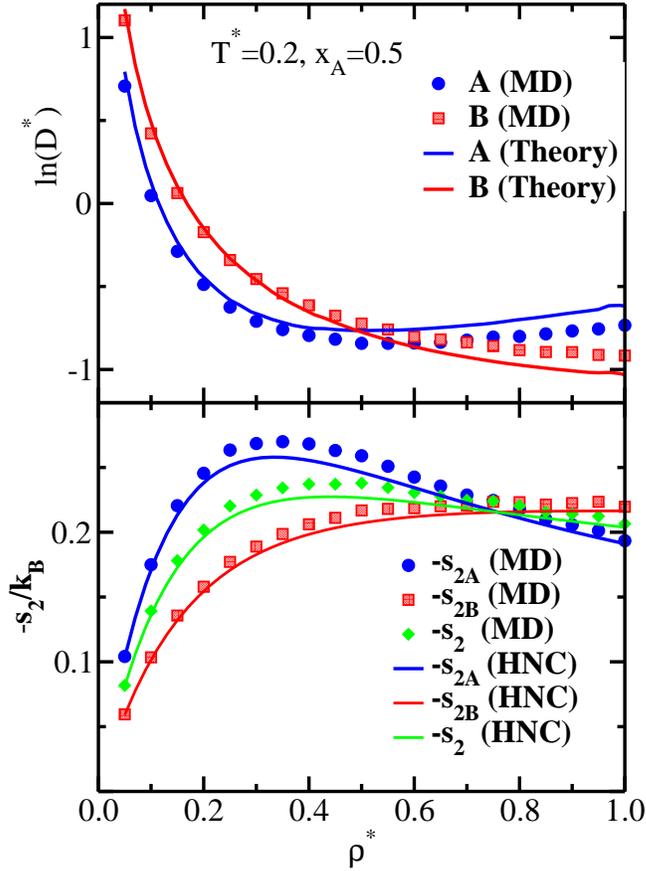}
\vspace{0.8cm} 
\caption{Upper panel: 
Simulation and theoretical results for the 
tracer diffusivities of a binary GC mixture as a function of fluid density
along the isotherm $T^*$=0.2 at the mole fraction $x_A$=0.5.
Lower panel: negative two-body contributions to the excess entropy of a GC
mixture.}
\label{figdsm}
\end{figure}

Finally, in connection to our treatment of 
time-dependent friction of GC fluid via short-time ansatz, we remark
that the short-time expansion coefficients for $\eta(t)$ have been 
reported\cite{sharma95} only up
to the term of order $t^4$. With this information, one would be able to
construct an ansatz of the form given in Eq.~(\ref{zetans}) only for the
time-dependent viscosity itself 
(or for its individual components defined in Eq.~(\ref{etasum})), but
not for its first order MF. Given that $\eta(t)$ at high densities displays a
pronounced biphasic behavior similar to that of $C_v(t)$, such approach clearly
would be inadequate, and we have not pursued it here. 

So far, our discussion was limited to a neat GC fluid. However, binary mixtures
of GC particles also exhibit structural and dynamical anomalies, both in terms
of density and mole fraction behavior. Both 
thermodynamic\cite{archer01} and dynamic\cite{pond09} properties of these
mixtures  have been studied extensively and the following observations were
made. For a binary mixture of particles of two different sizes, at
intermediate and high densities, the tracer diffusivity of the larger species
increases (and that of the smaller species decreases) either upon increasing the
density of the mixture (at fixed mole fraction) or upon increasing the mole 
fraction of larger species (at a given density). As in the case of a neat GC
fluid, a strong correlation was observed between these dynamical anomalies and
structural anomalies, as quantified by two-body contributions to excess entropy
of individual species. 

The aforementioned results for binary mixtures of GC particles were obtained
via MD simulations.\cite{pond09} In the present work, we attempt to calculate
the same structural and dynamical properties from a microscopic theory. 
We consider a two-component GC fluid containing species $A$ and $B$. The
pair interaction potential between particles has the form:
$\phi_{ij}(r)=\epsilon_{ij}\exp[-(r/\sigma_{ij})^2]$, where $i,j=A,B$.
In
order to be able to perform a direct comparison between theory and simulation,
we choose a  set of parameters used in previous studies:\cite{archer01,pond09}
$\sigma_{BB}=0.665\sigma_{AA}$, 
$\sigma_{AB}=(0.5[\sigma_{AA}^{2}+\sigma_{BB}^{2}])^{0.5}$,
$\epsilon_{AA}=\epsilon_{BB}$, $\epsilon_{AB}=0.944\epsilon_{AA}$. 
We also assume that the particles of two species have the same mass 
$m_A=m_B$. 

As in the case of a neat GC fluid, we compute $g_{ij}(r)$, 
the radial distribution functions of the mixture, from the integral equation
theory with HNC closure. The degree of translational structural order of the
mixture is quantified via two-body part of the excess entropy given 
by:\cite{pond09}
\be
s_2=\sum_{i}x_i s_{2i},
\label{s2mix}
\ee 
where $x_i$ is the mole fraction of species $i$, and 
$s_{2i}$, which characterizes the degree of pair structural ordering
surrounding particles of species $i$, is defined by:\cite{pond09}
\be
\frac{s_{2i}}{k_B}=-\sum_{j}2\pi x_j\rho\int_{0}^{\infty}dr r^2
\left[g_{ij}(r)\ln g_{ij}(r)-(g_{ij}(r)-1)\right]
\label{s2i}
\ee

In order to obtain the tracer diffusivities of the two species from MCT, one
needs to construct the corresponding time-dependent friction functions, 
$\zeta_{i}(t)$, which, in turn, requires the knowledge of the matrix of dynamic
and self-dynamic structure factors, $F_{ij}(k,t)$ and $F_{sij}(k,t)$. However,
computing $F_{AB}(k,t)$ from a continued fraction representation is highly
problematic,\cite{srinivas01} because its zero-time second order time
derivative is zero and the sign of the fourth derivative is oscillatory.
Instead, one could obtain $F_{ij}(k,t)$ from a time-dependent density
functional theory.\cite{srinivas01} However this approach emphasizes long-time
hydrodynamic-like behavior of dynamic structure factors, while our results for
a neat GC fluid indicate that transport anomalies can be explained on the basis
of the short-time behavior of TCFs. In particular the results for $D$ and
$C_v(t)$ of a neat GC fluid given by a short-time ansatz were for most part
comparable in accuracy to the MCT results. 
Accordingly, instead of constructing MCT
for a binary mixture, we have adopted a simpler approach by modeling 
$\zeta_i(t)$ via Eq.~(\ref{zetans}). This approach requires the knowledge of
the short-time expansion coefficients of velocity TCFs, $C_{vi}(t)$, up to
the term of order $t^6$. The corresponding expressions have been reported
earlier.\cite{sharma98} 

We present our results for tracer diffusivities in dimensionless form given by:
$D^{\ast}_{i}=D_i(m_A/\epsilon_{AA}\sigma_{AA}^{2})^{1/2}$, while dimensionless
 density of the mixture is defined by $\rho^*=\rho\sigma_{AA}^{3}$. 
Simulation\cite{pond09} and theoretical results for $D^{\ast}_{i}$ as a function
of fluid density along the isotherm $T^*=k_BT/\epsilon_{AA}$=0.2 
are given in the upper panel of Fig.~\ref{figdsm}  for the
mole fraction $x_A$=0.5. While the tracer diffusivity of the larger
species displays the same anomalous density behavior as the self-diffusion
coefficient of a neat GC fluid (i.e. passes through a minimum around 
$\rho^*\sim$0.4 and then increases with density), $D^{\ast}_{B}$ decreases
monotonically with $\rho^*$ throughout the entire density range studied. As a
result, the curves for the  two tracer diffusivities cross at a certain
intermediate density, beyond which the mobility of larger particles exceeds
that of the smaller particles. Theoretical results agree well with MD data
below the crossover density, while above this point ansatz overestimates the
tracer diffusivity of the larger species and underestimates that of the smaller
species. Still, all the trends in the density behavior of the two diffusivities
are reproduced correctly by the theory.  

\begin{figure}
\includegraphics[width=10cm,angle=0]{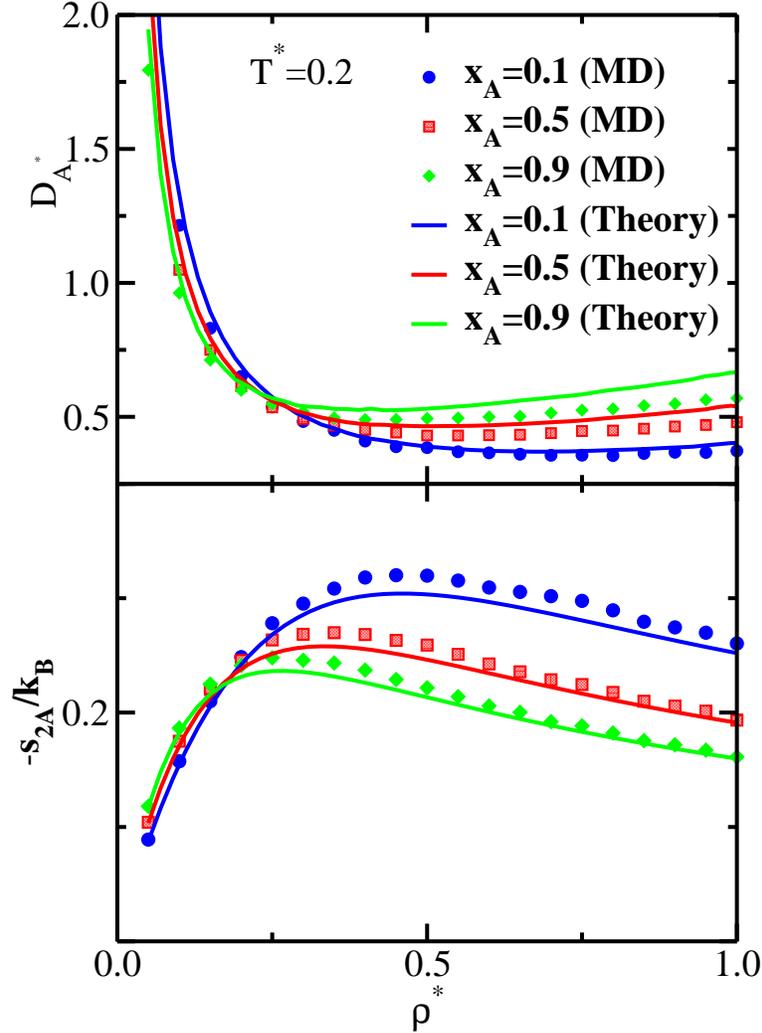}
\vspace{0.8cm} 
\caption{Upper panel: 
Simulation and theoretical results for the 
tracer diffusivity of the larger species of a binary GC mixture 
as a function of fluid density
along the isotherm $T^*$=0.2 for three values of the mole fraction of A.
Lower panel: negative two-body contributions to the excess entropy of the 
larger species of a binary GC mixture.}
\label{figdsa}
\end{figure}

Given that the only input into the theory is the structural information, i.e.  
the pair distribution functions $g_{ij}(r)$ entering the expressions for the
short-time expansion coefficients, one can expect a strong correlation
between structural and dynamical anomalies of a GC binary mixture, similar to
that observed for a neat fluid. Indeed, the lower panel of Fig.~\ref{figdsm}
displays the two-body contributions to the excess entropy given by 
Eqs.~(\ref{s2mix}) and (\ref{s2i}), and one sees that $-s_{2A}$ and $-s_{2B}$
show markedly different density behavior.  In analogy to the diffusivity,
$-s_{2A}$ as a function of $\rho^*$ follows the same pattern as $-s_2$ of a neat
fluid, namely, it initially grows with density, passes through a maximum, and
then decreases, reflecting more pronounced overlaps of larger particles at
higher 
densities.\cite{pond09} By contrast,  $-s_{2B}$ increases monotonically with
$\rho$, which again results in the crossing of the curves for the two species. 
Theoretical results for $s_{2i}$ agree well with the MD data,\cite{pond09}
indicating that HNC closure is reliable for calculating the structure of GC
mixtures.

\begin{figure}
\includegraphics[width=10cm,angle=0]{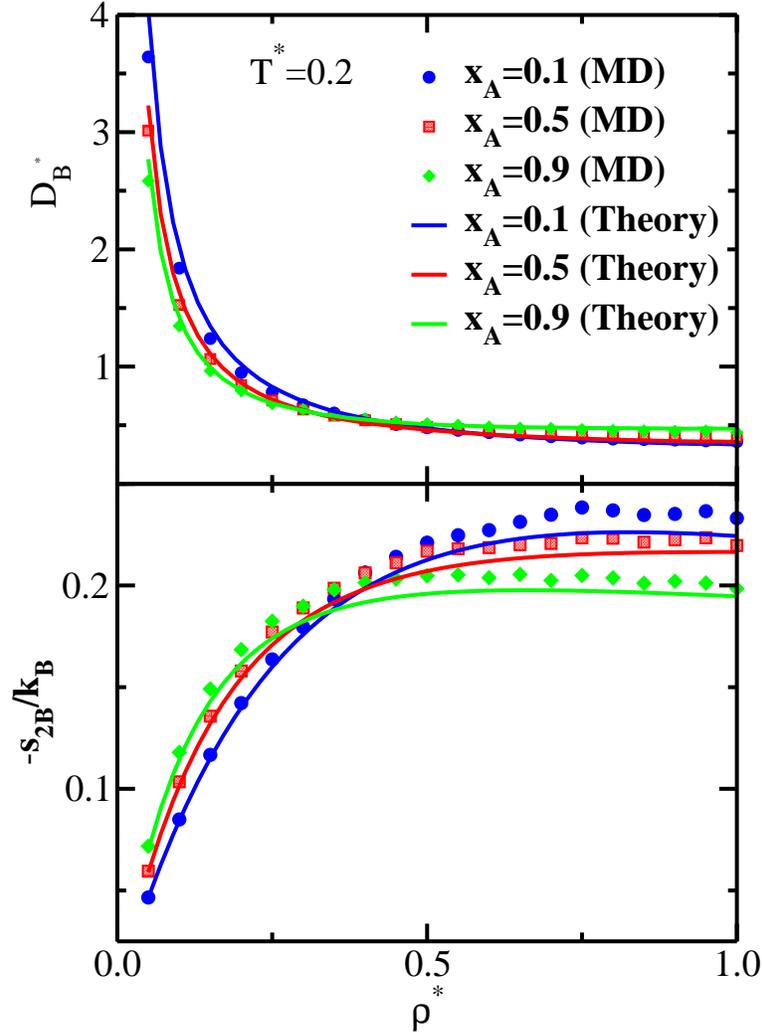}
\vspace{0.8cm} 
\caption{Upper panel: 
Simulation and theoretical results for the 
tracer diffusivity of the smaller species of a binary GC mixture 
as a function of fluid density
along the isotherm $T^*$=0.2 for three values of the mole fraction of A.
Lower panel: negative two-body contributions to the excess entropy of the 
smaller species of a binary GC mixture.}
\label{figdsb}
\end{figure}

Having analyzed the structural and dynamical anomalies of an equimolar GC
mixture, we now consider the effect of changing the mole fraction of larger
particles on the tracer diffusivities and structural order metrics of the two
species. Tracer diffusivity of the larger species is shown in the upper panel 
of Fig.~\ref{figdsa} as a function of fluid density along the isotherm $T^*$=0.2
for several values of $x_A$; the corresponding results for $-s_{2A}$ are shown
in the lower panel.  Tracer diffusivity and two-body excess entropy of the
smaller species at the same conditions are shown in the upper and lower panels
of Fig.~\ref{figdsb}, respectively. One sees that in the low density regime,
increase in $x_A$ results in decreasing tracer diffusivity and increasing
structural order metric for both species. As pointed out in the earlier
simulation study,\cite{pond09} this behavior can be expected, because increasing
mole fraction of the larger species results in higher packing fraction of the
fluid. By contrast, at higher densities, larger values of $x_A$ correspond to
more overlaps between particles, which translates into lower values of 
$-s_{2i}$ and higher values of $D_{i}$ for both species. While the trends in
the mole fraction behavior of  $-s_{2i}$ and  $D_{i}$ are the same for both
species, the switch from expected to anomalous dependence on $x_A$ occurs
earlier (i.e. at a lower value of bulk density) for the larger species.
Once again, theory successfully captures all the trends observed in simulations,
even though there are minor quantitative discrepancies, mostly at high 
densities. 

As a final remark on GC mixtures, we note that for a given density and
temperature one might expect anomalous (i.e. nonlinear) dependence of shear
viscosity on the mole fraction $x_A$.\cite{srinivas01} Given that our
short-time ansatz is inadequate for modeling time-dependent viscosity, we
cannot verify this conjecture presently. One possible approach to this problem
would be to obtain dynamic structure factors $F_{ij}(k,t)$ for the mixture from
a fully self-consistent MCT framework.\cite{flenner05} This will be the subject
of future investigation. 

\section{Conclusion}
\label{sc4}

In summary, we have reported a theoretical study of structural and 
dynamical anomalies of a neat GC fluid and GC binary mixture. 
As has been discussed previously in the
literature,\cite{krekelberg09,krekelberg09b,pond09}  
essentially all the
observed anomalies can be traced back to the fact that GC potential is bounded
and, as such, allows interparticle overlap, which becomes more and more
prominent at higher densities. We have employed integral equation theory to
compute the GC liquid structure and both MCT approach and short-time ansatz for
MFs to obtain its transport coefficients. Theory was successful in reproducing
anomalous density behavior of both diffusion coefficient and shear viscosity,
where the former increases and the latter decreases with density. One major
difference between the two is that viscosity displays anomalous behavior over
a limited density range only, following which it again increases with density,
while diffusion anomaly persists over the entire density range studied. This
fact was rationalized by noting that viscosity can be split into potential and
kinetic contributions, with the former displaying anomalous density
dependence and the latter behaving normally. With increasing density and/or
temperature, the relative importance of the
kinetic term increases, and viscosity reverts from anomalous to normal density
behavior. 

A deep connection between structural and dynamical anomalies uncovered in
earlier studies\cite{krekelberg09,krekelberg09b,pond09} has been reconfirmed by
noting that both density and mole fraction anomalous behavior of diffusion could
be described via a short-time ansatz for a time-dependent friction, with the
latter constructed from structural data only. One remaining open question
concerns the mole fraction dependence of shear viscosity of a GC mixture, it
will be the subject of future research.

\section{Acknowledgment}
The authors would like to thank 
Profs.~T.~M.~Truskett and H.-O. May and 
Drs.~W.~P.~Krekelberg and M.~J.~Pond for sending the simulation data. 
One of us (S. A. E.), acknowledges support from the Alexander von Humboldt
foundation, Germany.


\end{document}